# Effect of temperature on 2D terahertz plasmons and electron effective mass in AlGaN/GaN

M. Dub[1,2*], P. Sai[1,2], D. Yavorskiy[1,2], Y. Ivonyak[1,2], P. Prystawko[2], R. Kucharski[2],

G. Cywiński[1,2], W. Knap[1,2], S. Rumyantsev[2]

[1]CENTERA, CEZAMAT, Warsaw University of Technology, 02-822 Warsaw, Poland

[2] Institute of High Pressure Physics PAS, 01-142 Warsaw, Poland

*Correspondence: maksimdub19f94@gmail.com

## Abstract

The effect of temperature on two-dimensional plasmons in large-area AlGaN/GaN plasmonic crystals was studied experimentally. With the temperature increase the resonant plasmon frequency redshifts due to the strong temperature dependence of the electron effective mass and electron concentration under open to the environment surface of AlGaN. The temperature dependence of electron effective mass is confirmed by the cyclotron resonance measurements.

1. Introduction

Collective oscillations of the charge carriers in semiconductors (plasmons) have been of great interest for over 50 years. Two-dimensional (2D) plasmons excited in semiconductor quantum wells (QWs) or 2D materials, like graphene are of special interest because in the case of high 2D charge carrier density ($\sim 10^{13}$ cm$^{-2}$) and for the devices with lateral dimensions in the micrometer range their frequency lies in the terahertz (THz) frequency band.

One of the ways to study 2D plasmons is to measure the optical properties of semiconductor structures of large area with a periodic charge density profile. This kind of structures, known as plasmonic crystals (PCs) can be formed, for example, by integration of metal gratings with inversion layers in semiconductors, heterostructures with QWs, and 2D materials. Theoretical and experimental studies of these grating-gate structures started in the 1980s using Si inversion layers and currently continues using AlGaAs, AlGaN and graphene–based devices (see ref. [1] and refs. therein). It was shown experimentally and by numerical simulations that periodic structures behave as PCs and their properties depend on carriers' concentration profile and PCs period [1,2]. Although all details of PCs behavior can be obtained only by rigorous electrodynamic simulations, the basic properties of plasmons can be found from an analytical model that describes well the experiments [3,4]. The wide spectrum of PC properties and ease of fabrication make PCs promising for devices like detectors, tunable filters, amplifiers, and generators operating at THz frequencies. It was already demonstrated experimentally that PCs can be used as different kinds of passive devices, like detectors and filters [5-11].

Depending on the gate voltage, two types of plasmon modes in PCs can be distinguished. When the gate voltage is such that the 2D concentrations in gated and ungated parts of a PC are comparable, the delocalized mode can be observed [1]. In this case, the plasmon properties are determined by the interacting gated and ungated parts of the PC. When the concentration under the gate tends to zero, ungated parts play a major role and so called localized mode of the PC is formed [1]. The localized mode (also called unscreened or ungated mode) has an advantage of higher frequency for the same characteristic dimensions. The frequency of both the delocalized and localized modes can be adjusted by the gate voltage but the physical phenomena responsible for the tuning are different. As was shown in refs. [1,12] the frequency of delocalized modes is tuned by carrier density under the gated regions of the PC, whereas the localized mode tuning is related to the effect of shrinking length of the ungated parts (side gate effect) or by additional back gate.

Here we studied the temperature dependence of both delocalized and localized plasmon modes properties. It was demonstrated already in several publications that the plasmon resonant frequency in the delocalized mode depends on temperature and this dependence was explained by the strong temperature dependence of the electron effective mass. Since this is not a trivial and expected conclusion, we repeated those experiments and analyzed possible mechanisms of the plasmon frequency temperature dependence. We found that two parameters contribute to the temperature dependence of the plasmon frequency: electron concentration in the ungated parts of the PC and electron effective mass. We performed the cyclotron resonance measurements which confirmed the strong temperature dependence of the effective mass.

We discuss a few different possible physical mechanisms of temperature dependence of the effective mass in AlGaN/GaN heterostructures and show that until now there exist no any plausible explanation of this effect. Therefore, our results state an important fundamental question for future studies.

2. **Experimental details**

Three distinct types of AlGaN/GaN heterostructures were employed to fabricate large-area PCs: commercially available heterostructures on SiC substrates from "SweGaN" (Linköping, Sweden [13]), the heterostructures fabricated at the Institute of High Pressure Physics (IHPP) on SiC substrates, and heterostructures also fabricated at IHPP on native semi-insulating GaN:Mn substrate grown by ammonothermal method [14]. All structures with 2D electron gas (2DEG) at the AlGaN/GaN interface were grown by metal-organic vapor phase epitaxy (MOVPE). Table I shows the main parameters and layers' stacks description of these heterostructures.

Two different types of large PCs, which are schematically shown in Figs. 1(a)–1(b), were fabricated and studied. Figures 1(c)-1(d) show the optical microscope images of the corresponding PCs.

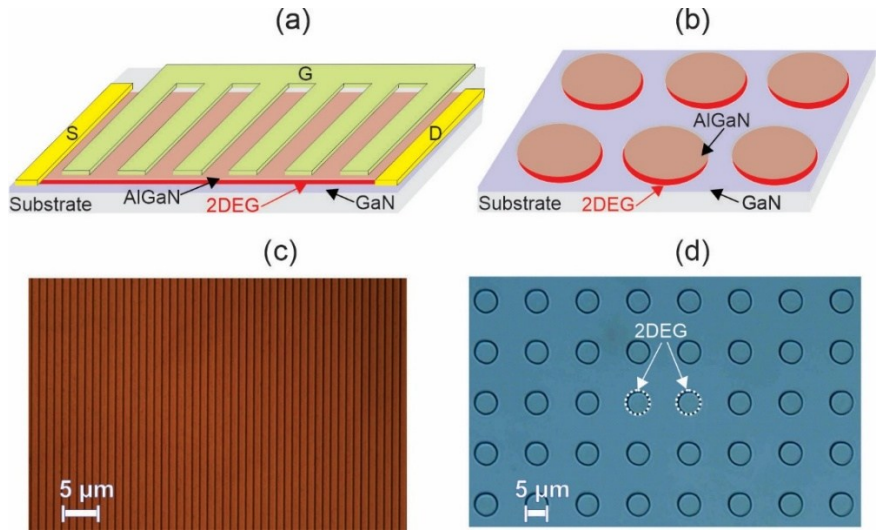

FIG. 1. (a)-(b) Schematics and (c)-(d) optical microscope images of the studied devices. (a),(c) Grating gate FET. (c), (d) periodic structure of discs with 2DEG and 2DEG removed between them by RIE.

The commercial laser writer system (Microtech, Palermo, Italy) equipped with a 405 nm GaN laser source was used to define the elements of micrometer resolution, like contact pads and the elements for the PCs in Fig. 1(a), 1(b). Electron beam lithography (EBL) was employed to create a large-area gratings with sub-micrometer resolution for the PCs shown in Fig. 1(a).

For the ohmic contacts, the Ti/Au/Ni/Au (150/1000/400/500 Å) metal stacks were deposited and subsequently annealed at 800 °C in a nitrogen atmosphere for 1 minute. To fabricate the Schottky barrier gates, Ni/Au (150/350 Å) metal stack was used (see ref.[1] for more details). The main parameters of the fabricated PCs are shown in Table II.

The transmittance spectra of the plasmonic crystal were studied using a Fourier-transform infrared (FTIR) vacuum spectrometer (Vertex 80v, Bruker, Billerica, MA, USA) equipped with a continuous-flow liquid helium or nitrogen cryostat (Optistat CF-V, Oxford Instruments, Abingdon, UK). The temperature controller (MERCURY-ITC-1, Oxford Instruments, Abingdon, UK) was synchronized with an electrical heater and a temperature sensor placed near the sample on the cold finger inside the cryostat. During temperature-dependent measurements, an additional delay of 5 minutes was applied before the start of the measurement to ensure thermal equilibrium and minimize temperature fluctuations. At each temperature point, the measurement was performed twice to verify temperature stabilization by ensuring the absence of changes in the spectral characteristics. Fabry-Pérot oscillations from the sample substrate were eliminated using Fourier transformation with a step size of 4 cm$^{-1}$ (0.12 THz). The spectrometer was equipped with a cryogenically cooled silicon bolometer (General Purpose 4.2 K Bolometer System, IRLabs, Tucson, AZ, USA), a solid-state silicon beam splitter, and an external water-cooled high-power Hg-arc lamp as the radiation source (Bruker, Optik GmbH, Billerica, MA, USA). To prevent oversaturation of the Si-bolometer, a 2 mm aperture was positioned directly after the sample, restricting electromagnetic radiation transmission to the size of the plasmonic crystal active region. Electrical control of the plasmonic crystal and DC measurements within the FTIR spectrometer were performed using a Keysight B2902A Precision Source/Measure Unit.

Cyclotron resonance experiments were performed using an untreated AlGaN/GaN structure positioned within a variable temperature insert at the center of a superconducting magnet, ensuring that the 2DEG was oriented perpendicular to the magnetic field. The sample temperature was controlled between 10 K and 290 K via a programmable PID temperature controller. Monochromatic THz radiation at 0.66 THz was transmitted to the sample via a cylindrical stainless steel waveguide of 8 mm internal diameter, with its termination positioned 5 mm from the sample surface. The transmitted radiation was then collected through a waveguide located outside the cryostat, and its intensity was measured using a pyroelectric detector in combination with lock-in amplification.

Table I. AlGaN/GaN heterostructures used to fabricate PCs: layers and thicknesses (from top to bottom).

| Heterostructure stack | SL 3665 (SweGaN) | Hx5612 (IHPP) | Hx6341A (IHPP) |
|---|---|---|---|
| Cap layer | 2.4 nm GaN | 2 nm GaN | 2 nm GaN |
| Barrier layer | 20.5 nm $Al_{0.25}Ga_{0.75}N$ | 20 nm $Al_{0.28}Ga_{0.72}N$ | 12.5 nm $Al_{0.24}Ga_{0.76}N$ |
| Spacer layer | - | 0.88 nm AlN | - |
| Channel layer | 155 nm GaN | 3 μm GaN | 0.5 μm GaN |
| Buffer layer | - | - | 1.0 μm GaN:C |
| Nucleation layer | 62 nm AlN | 40 nm AlN | - |
| Substrate | 500 μm SiC | 500 μm SiC | 400 μm Ammono GaN |

Table II. Main parameters of the studied PCs.

| sample ID | Cavity lengths, gated/ungated $L_G/L_{UG}$, μm | Number of cavities: gated/ungated | Active area, mm$^2$ | Total channel width, mm | Type of heterostructure and manufacturer |
|---|---|---|---|---|---|
| S1 | 1.6/0.4 | 850/851 | 2.89 | 1.7 | Hx5612 (IHPP) |
| S2 | 1.2/03 | 1133/1134 | 2.89 | 1.7 | Hx5612 (IHPP) |
| S3 | 1.6/04 | 850/851 | 2.89 | 1.7 | Hx6341A (IHPP) |
| S7 | 0.9/06 | 550/550 | 2.89 | 1.7 | SL 3665 (SweGaN) |
| L1 | 0/ diameter 5 | 0/40000 | 4 | | SL 3665 (SweGaN) |
| L2 | 0/ diameter 5 | 0/40000 | 4 | | Hx5612 (IHPP) |

Figure 2 shows examples of the transfer current-voltage characteristics for one of the grating gate PC (S7) at different temperatures. As seen, despite the very high gate area the gate voltage, $V_G$, effectively controls the drain current, $I_D$, providing the virtually zero subthreshold current at $V_G<V_{th}$. The threshold voltage, $V_{th}$, can be estimated from the linear approximation of the current-voltage characteristics as shown in Fig. 2. It is important that the threshold voltage does not depend on temperature, which is an indication that the electron concentration in the channel does not depend on temperature as well, and it can be found as

$$n_s = \frac{C(V_G-V_{th})}{e}, \qquad (1)$$

where $n_s$ is the 2DEG concentration, $C$ is the gate capacitance per unit area, and $e$ is the elementary charge.

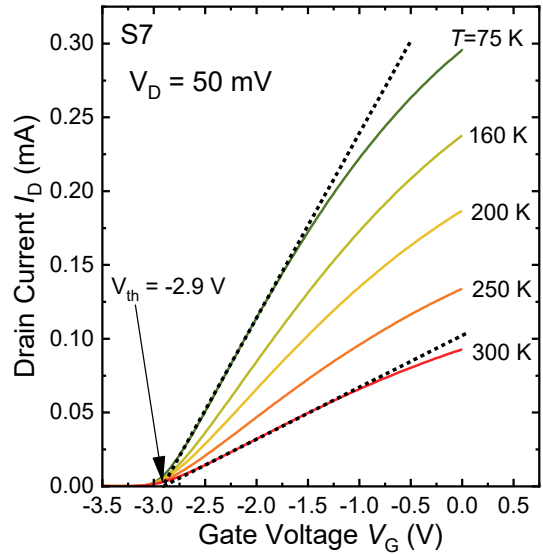

FIG. 2. Transfer current-voltage characteristics of the grating gate S7 PC at different temperatures. Dashed lines show the linear approximation and provide the threshold voltage $V_{th}$ at the intersection with the horizontal axis ($I_D=0$).

## 3. Results and discussions

The effect of temperature on the resonant frequency of 2D delocalized plasmons in AlGaN/GaN QWs was found in a few earlier publications. First, it can be seen in the experimental data of ref. [15], where it was not discussed yet. A more detailed study of the plasmon temperature dependence for the first harmonic of the delocalized mode in AlGaN/GaN grating-gate structures is presented in ref.[16].

Figure 3 shows the transmittance spectra for S7 PC at different temperatures and at $V_G=0$. As seen, the plasmon frequency redshifts with the temperature increase for both plasmon harmonics (see Fig. I in the supplementary materials for the transmittance spectra of S2 and S3 PCs at different temperatures).

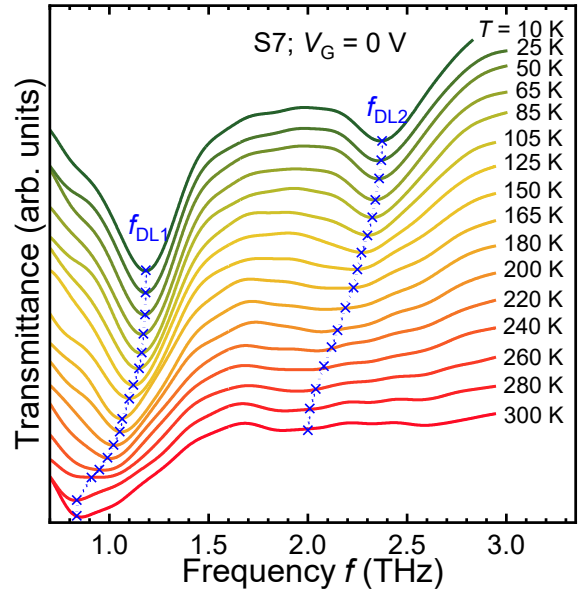

FIG. 3. Transmittance spectra for the first and second harmonics of delocalized plasmons at different temperatures for S7 PC. All spectra are shown with vertical shifts for better visibility. Blue crosses show the position of the plasmon resonant frequencies. Dashed lines are guides for the eye.

For the single cavity approach, the frequency of 2D plasmons is given by [17,18]:

$$f_p = \frac{1}{2\pi}\sqrt{\frac{2\pi e^2 n_s q}{m^* \varepsilon_{eff}}}, \qquad (2)$$

where $q$ is the wave vector for the first and higher harmonics, $m^*$ is the effective mass, and $\varepsilon_{eff}$ is the effective dielectric permittivity, which depends on the wave vector, the materials' dielectric permittivity $\varepsilon$, and the structure geometry.

For the screened (gated) plasmons eq. (2) is modified as:

$$f_p = \frac{q}{2\pi} s_g = \frac{q}{2\pi}\sqrt{\frac{4\pi e^2 n_s d}{m^* \varepsilon}}, \qquad (3)$$

where $s_g$ is the gated plasmon velocity and $d$ is the barrier thickness.

However, eqs.(2) and (3) do not yield the correct frequency of plasmons in PCs because PCs represent periodic structures with gated and ungated interacting cavities with different plasmon velocities [1].

It was shown in ref. [4] that the analytical model of the plasmonic crystal, proposed in refs. [1,2], describes well the PCs' basic behavior. In this model, the resonant frequencies (delocalized mode) can be found as the solutions of the equation:

$$\cos\frac{\omega_p L_1}{s_1}\cos\frac{\omega_p L_2}{s_2} - \frac{s_1^2 + s_2^2}{2 s_1 s_2}\sin\frac{\omega_p L_1}{s_1}\sin\frac{\omega_p L_2}{s_2} = 1, \qquad (4)$$

where $L_1$ and $L_2$ are the lengths of the plasmonic cavities, and $s_1$ and $s_2$ are the corresponding plasma wave velocities. Taking $L_1$, $s_1$, and $L_2$, $s_2$ as characteristics of the gated and ungated parts of the plasmonic crystal, respectively, we can estimate the plasma wave resonant frequencies $f_{DL}=\omega_p/2\pi$. The gated, $s_g$, and ungated, $s_{ug}$, plasmon velocity can be found as:

$$s_G = \sqrt{\frac{4\pi e^2 n_{sg} d}{m^* \varepsilon}} \qquad s_{UG} = \sqrt{\frac{4 e^2 n_{sug} L_2}{(\varepsilon+1) m^*}}, \qquad (5)$$

where $n_{sg}$ and $n_{sug}$ are the 2DEG concentrations in the gated and ungated parts of the PC. These velocities determine the plasmon frequency and its dependence on electron concentration, effective mass, and dielectric permittivity, which all might be temperature-dependent.

As seen in Fig. 2, the threshold voltage of these grating-gate structures either does not depend on temperature or even slightly increases with the temperature increase. Therefore, we can conclude that concentration, $n_{sg}$, either *does not depend* on temperature or only slightly increases with the temperature increase. Therefore, the temperature dependence of concentration $n_{sg}$ cannot explain the decrease of the plasmon frequency with the temperature increase.

It is known that dielectric permittivity, $\varepsilon$, is only weakly temperature dependent in semiconductors. For example, for relatively wide-band gap semiconductors, GaAs and GaP dielectric permittivity increases only ~3-4% when temperature is changed from 4K to 300K [19]. We are not aware of the publications where the temperature dependence of GaN low-frequency dielectric permittivity was studied. However, the temperature dependence of GaN optical dielectric permittivity measured at high temperatures is also weak [20]. The same as in other semiconductors, the change in the GaN dielectric permittivity does not exceed a few percent over a 300 K temperature range. Having in mind that the plasmon frequency is proportional to the square root of $\varepsilon$, we can conclude that the temperature dependence of dielectric permittivity *gives a negligible correction* to the plasmon frequency.

One more mechanism of the plasmon properties temperature dependence is the effect of the electron concentration in the ungated parts of the PC. The plasmon resonant frequency in PC depends on the plasmon velocities in two parts of the structure with two different electron concentrations. We already discussed that the concentration under the gate can be easily estimated from the transistor threshold voltage and showed that it does not depend on temperature. However, this might not be the case for the ungated region, which is exposed to the environment. Due to the effect of temperature on the surface states, the electron concentration in the ungated parts might depend on temperature. In order to check it, we measured the plasmon transmittance spectra for the localized modes when the concentration under the gate is virtually zero (large negative gate voltage). In this modes, the plasmon frequency is determined only by the concentration in the ungated part of the PC and by its geometry. The temperature dependences of the transmittance spectra in the L1 and L2 PCs were also measured. These structures do not have metal gates and the plasmon frequencies are determined by the properties of ungated islands and their period.

Figure 4 shows the localized plasmon transmittance spectra at different temperatures during the heating cycle from 70 K to 300 K for S7 and L1 PCs. The temperature dependences of the localized plasmon frequency during both heating and cooling cycles are shown in Fig. 5 (see Fig. II, and III in the supplementary materials for the transmittance spectra of L1 and S7 PCs measured with the smaller temperature step and for both heating and cooling cycles). As seen the frequency dependences demonstrate hysteresis behavior and non-monotonic character. Particularly, it is seen that frequency sharply increases at $T = 180$-$200$ K during the heating cycle, indicating the change of the surface charge and the corresponding change of 2DEG concentration.

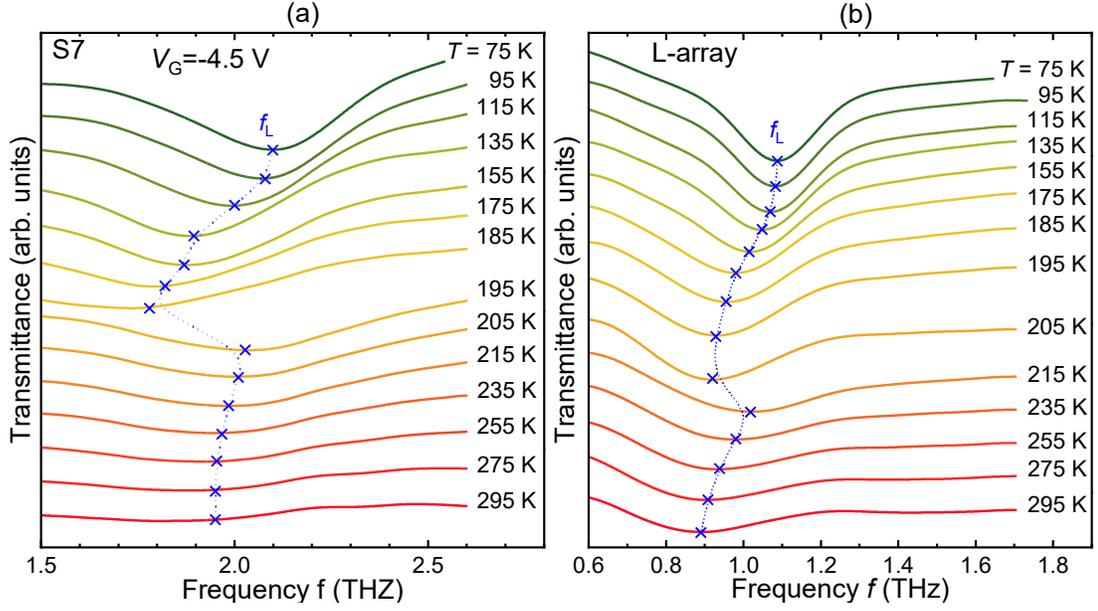

FIG. 4. (a) Transmittance spectra at different temperatures during the heating cycle for the localized plasmons in S7 and (b) L1 PCs. Blue crosses show the position of the plasmon resonant frequencies. Dashed lines are guides for the eye.

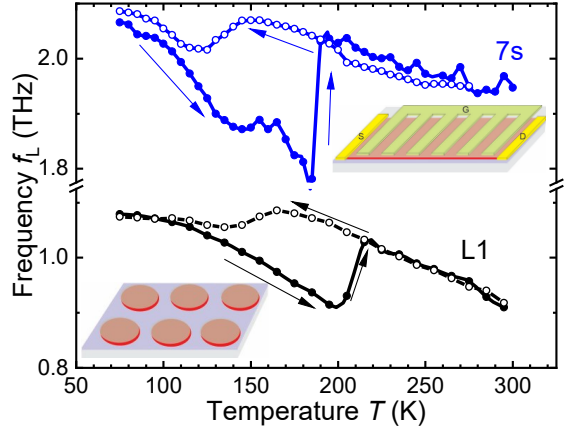

FIG. 5. Temperature dependences of the plasmon frequencies during the heating and cooling cycles for the localized plasmon mode in S7 (blue circles) and L1 samples (black circles). Insets show schematically the corresponding structures.

To further confirm the importance of the surface states we measured transmittance spectra under ultraviolet (UV) illumination. We used unfocused light from the light-emitting diode with $\lambda$ = 270 nm peak wavelength (ProLight Opto PB2D-1CLA-TC). The quanta energy for this wavelength is above the Al$_{0.25}$Ga$_{0.75}$N barrier layer bandgap of at 70 K. Therefore, we can expect effective generation of electron-hole pairs in AlGaN and an increase of the electrons' concentration in the QW. The estimate for the UV optical power using the nominal power of LED and the ratio of the PC area to the area of the actual light spot yields a small value of the incoming radiation power $P \approx 50$ μW.

Figure IV(a) and (b) in the supplementary materials show the transmittance spectra at different temperatures in the dark and under UV illumination for L2 PC. Figure IV(c) illustrates the effect of UV light on the plasmon resonant frequency for the L2 PC. As seen, due to the increase of the electron concentration in the QW, the plasmon frequency increases even under this weak illumination. The effect of light on the plasmon frequency is due to the two main processes: direct generation of electrons in GaN and change of the surface states under UV illumination. As seen in Fig. IV (d), UV light not only increases the plasmon frequency but also reduces the frequency jump at $T \approx 220$ K. This is a result of the surface state "cleaning" by UV illumination. It is also an indication of the importance of the surface states in defining the 2D concentration in the ungated channel and the corresponding plasmon frequency. Therefore, we can conclude that, indeed temperature dependence of the electron concentration in the ungated parts affects the localized plasmon frequency.

In general, this effect can affect the frequency of delocalized plasmons as well, because their frequency is determined by the plasmon velocities in both parts of the structure, i.e. gated and ungated.

The temperature dependence of the electron effective mass can also contribute to the plasmon resonant frequency temperature dependence, as was proposed in the paper by Pashnev et al. [16]. Figure 6 shows the temperature dependence of the $(f_{DL70}/f_{DL})^2$ parameter, calculated for the first harmonic of delocalized plasmons in S7 and S2 PCs, and for localized plasmons in PCS S7, L1, L2 ($f_{DL}$ and $f_{DL70}$ are the plasmon frequencies at given temperature and at $T=70K$, respectively). This parameter shows the relative increase of the effective mass assuming other parameters to be temperature independent. Right axis shows the effective mass calculated assuming the effective mass at $T = 70$ K $m^{*}_{70}/m_e=0.23$ ($m_e$ is the free electron mass).

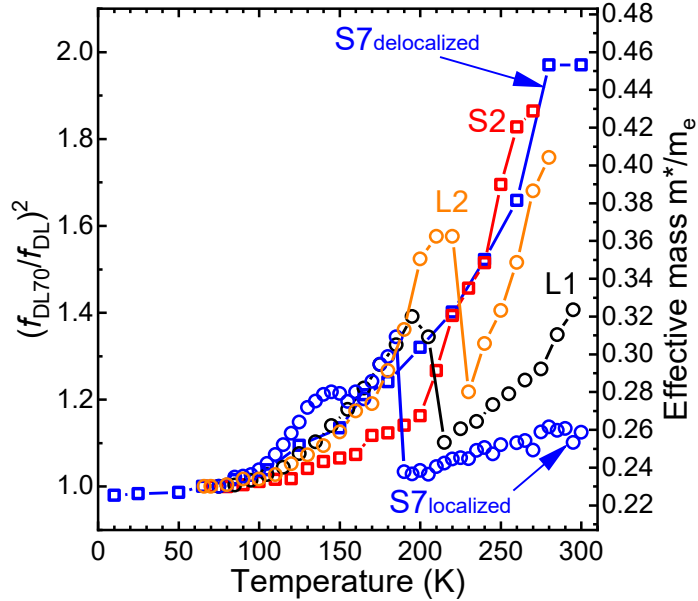

FIG. 6. Temperature dependence of the $(f_{DL70}/f_{DL})^2$ parameter for the delocalized mode of the plasmon resonance for S7, S2 PCs ($V_G=0$) and localized plasmons in PCs S7 ($V_G= -4.5$ V), L1, L2. Right axes shows the effective mass calculated as $\frac{m^*}{m_e} = \left(\frac{f_{DL70}}{f_{DL}}\right)^2 \left(\frac{m^*_{70}}{m_e}\right)$ assuming the effective mass $m^{*}_{70}/m_e=0.23$ at $T = 70$ K.

As seen in Fig.6, the experimental data for the temperature dependence of the plasmon resonant frequency is characterized by high dispersion (see the supplementary materials for the results on higher harmonics resonant frequency temperature dependence). The same conclusion can be made analyzing the temperature dependence of the higher plasmon harmonics (see the supplementary materials for more details). Since the assumption that the effective mass is different in different structures looks unrealistic, we have to conclude that the surface states in the ungated parts of the PCs strongly affect the plasmon frequency. Ungated parts of the PCs are open to the environment and are subject to different humidity, illumination and temperature providing high dispersion of the experimental data between different samples.

However, the temperature dependence of the electron effective mass can still contribute to the temperature dependence of the plasmon frequency. The most straightforward way to find the effective mass in semiconductors is the cyclotron resonance (CR) measurements. This method is usually used at low temperatures when the mobility is high enough to observe the spectral line of the CR absorption. It is important that cyclotron frequency does not depend on carriers' density. Figure 8(a) shows the transmission of 0.66 THz radiation at different temperatures of "SweGaN" AlGaN/GaN plane wafer as a function of the magnetic field. As seen, the absorption lines become wider and shift to higher magnetic fields with the temperature increase. Widening of the spectral line is a result of the mobility decrease with the temperature increase, i.e. due to the increasing losses. In general, losses may lead to the shift of resonance position as well. However, the calculation of the resonance line shape taking into account the losses [21] shows that losses lead to the shift of the line to the lower magnetic fields, which is the opposite trend to what is seen in Fig. 8(a). Although the spectral lines in Fig. 8(a) are significantly broadened at $T > 200$ K the shift of the spectra minima to the higher magnetic fields still can be identified which is an indication of the effective mass increase. Fig. 8(b) shows the temperature dependence of the effective mass extracted from the CR measurements as:

$$m^*/m_e = \frac{eB_{min}}{2\pi f_c m_e}, \qquad (2)$$

where $B_{min}$ is the magnetic field corresponding to the minimum of transmission (CR), $f_c$ = 0.66 THz. Due to the significant broadening of the spectral lines at $T > 200$ K the accuracy of the effective mass estimate at high temperature is low. However, the increase of the effective mass with the temperature increase can still be justified. Interestingly, CR measurements in AlGaAs/GaAs QWs in ref.[22] also showed the substantial increase in the effective mass with the temperature increase.

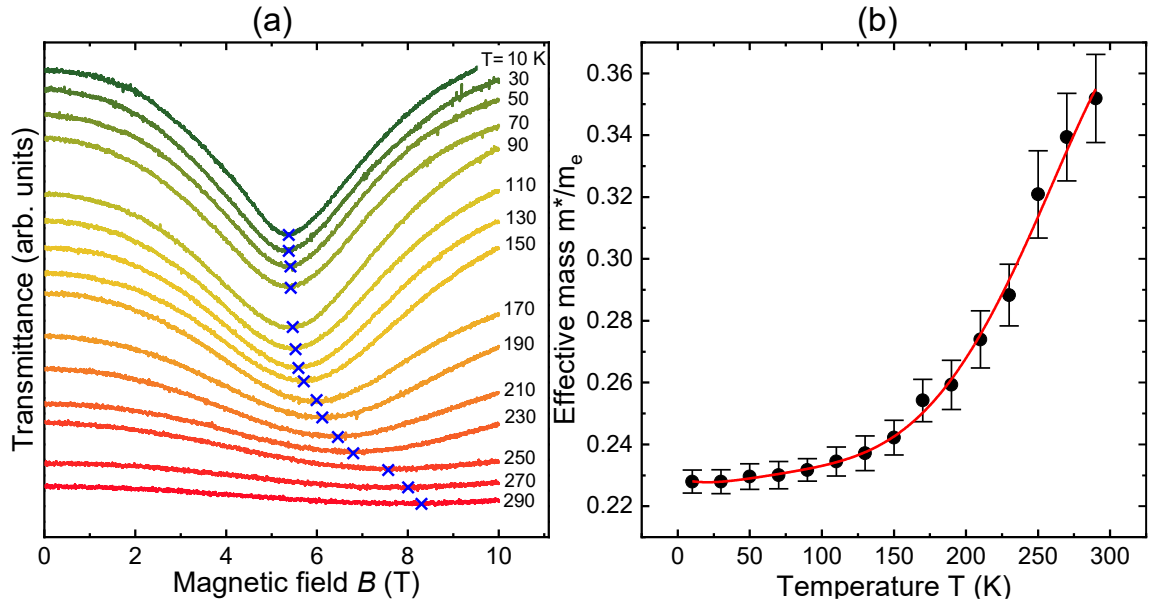

FIG. 8. (a) Cyclotron resonance transmittance of $f_c$=0.66 THz radiation as a function of the magnetic field and (b) electron effective mass extracted from the position of transmittance minima in the (a) panel. ("SweGaN" heterostructure).

The temperature dependence of the effective mass in GaN and AlGaN/GaN heterostructures was discussed in a few recent papers. The temperature dependence of the effective mass was evaluated by measuring Drude conductivity in the plane AlGaN/GaN heterostructures [23] and by THz optical Hall effect in thick GaN epitaxial layers and AlGaN/GaN heterostructures [24-27]. In the majority of the experiments, the increase of the effective mass with temperature was observed. The effect was not observed in epitaxial layers of AlGaN/GaN heterostructures with AlN spacer [22,28]. Indeed, hybridization of the envelope wave functions of electrons in the conduction band of GaN may lead to the effective mass increase, and the AlN spacer may prevent the effect. However, a simple estimate taking into account the effective masses of GaN and AlN and assuming linear dependence of the effective mass on the Al mole fraction in AlGaN shows that this effect is not significant. Later on, the effect of AlN layer between the barrier layer and 2D channel on the temperature dependence of the effective mass was not confirmed [21,23]. In ref. [29] the temperature dependence of plasmon resonances in AlGaN/GaN heterostructures was analyzed numerically. It was shown that inelastic scattering on optical phonons at $T > 70$ K leads to the redshift of the plasmon resonances.

There is some other evidence that optical phonons inelastic scattering affects the measured effective mass. First, as seen in Fig. 6 (see also [16,27]), just at $T > 100$ K when the optical phonon scattering becomes dominant, the major increase in the effective mass is seen experimentally. Second, in ref. [26] it was demonstrated that the only sample where the effective mass increase was not observed was the heavily doped GaN layer, where mobility is low and is determined not by the optical phonons but by the impurity scattering. This is an agreement with older results of the effective mass measurements in GaN at room temperature. In refs. [30,31], the effective mass was estimated from the optical measurements. It was found that the results of the measurements of heavily doped low mobility samples agree well with the value of the effective mass at $T = 300$ K of $m^*/m_e$=0.2-0.25. It was also mentioned in [30] that the results for low-doped sample do not agree with this value of the effective mass which should be higher than the mentioned value. In a more recent paper [32] the reflectivity and Hall effect measurements of the effective mass confirmed the value of $m^*/m_e = 0.20$-$0.24$ at $T = 300$ K but again for a heavily doped sample.

Indeed, it is well known that electron-phonon coupling due to the polaron effect increases the effective mass. However, the estimate for the polaron mass $m_p \approx m\left(1 + \frac{\alpha_e}{6}\right)$ yields an increase of only ~8% (see also [26]), which is significantly smaller than shown in Fig. 8(b) (here $\alpha_e = 0.49$ is the Fröhlich coupling constant [30]).

Therefore, we believe that although the unusually strong temperature dependence of the effective mass in the AlGaN/GaN system is experimentally observed, the physical mechanism of it is still unknown. Along with the temperature dependence of the surface state in the ungated parts of the PCs it defines the temperature dependence of the plasmon resonant frequency.

## 4. Conclusions

The temperature dependence of both delocalized and localized modes of plasmons in large area AlGaN/GaN PCs was studied experimentally. With the temperature increase the resonant frequency of both plasmon modes decreases. The analysis shows that this temperature dependence of plasmon frequency is due to the temperature dependence of the electron concentration in the ungated parts of the PCs and due to the temperature dependence of the electron effective mass.

The temperature dependence of the plasmon frequency is characterized by high dispersion of the experimental data for different PCs due to uncertainty of the surface states on the open surface of AlGaN. This also make impossible the estimate of the electron effective mass from plasmon resonant frequency measurements, as was proposed earlier. Nevertheless, we found that the electron effective mass indeed depends strongly on temperature, which we confirmed by the cyclotron resonance measurements.

Although the temperature dependence of the electron effective mass has been reported already by other methods, the physical mechanism of it is still unknown. Therefore, we believe that unusually strong temperature dependence of the effective mass in the AlGaN/GaN system is an important fundamental question for future studies.

SUPPLEMENTARY MATERIAL

See the supplementary material for additional information and figures.

ACKNOWLEDGMENTS

The work was supported by the European Union through the ERC-ADVANCED grant TERAPLASM (No. 101053716). Views and opinions expressed are, however, those of the author(s) only and do not necessarily reflect those of the European Union or the European Research Council Executive Agency. Neither the European Union nor the granting authority can be held responsible for them. We also acknowledge the support of "Center for Terahertz Research and Applications(CENTERA2)" project (FENG.02.01-IP.05-T004/23) carried out within the "International Research Agendas" program of the Foundation for Polish Science, co-financed by the European Union under European Funds for a Smart Economy Programme. We gratefully acknowledge Professor J. Łusakowski for his support in the cyclotron resonance measurements.

AUTHOR DECLARATIONS

Conflict of Interest

The authors declare no conflicts of interest.

Author Contributions

M. Dub: Conceptualization (lead); Data curation (lead); Investigation (lead); Methodology (equal); Project administration (lead); Writing– original draft (lead); Writing– review & editing (equal). P. Sai: Investigation (equal); Methodology (equal); Writing– review & editing (lead). D. Yavorskiy: Investigation (equal); Methodology (equal); Writing– review & editing (equal). Y. Ivonyak: Resources (equal); Investigation (equal). P. Prystawko: Resources (equal); Writing– review & editing (equal). R. Kucharski: Resources (equal); Validation (equal). G.Cywiński; Resources (equal); Writing– review & editing (equal). W. Knap; Conceptualization (equal); Supervision (lead); Validation (lead); Funding acquisition (lead); Writing– review & editing (equal).S. Rumyantsev: Conceptualization (lead); Data curation (lead); Methodology (lead); Project administration (lead); Writing– original draft (lead); Writing– review & editing (equal); Supervision (equal); Validation (equal).

# Effect of temperature on 2D terahertz plasmons and electron effective mass in AlGaN/GaN

## (supplemental materials)

M. Dub[1,2*], P. Sai[1,2], D. Yavorskiy[1,2], Y. Ivonyak[1,2], P. Prystawko[2], R. Kucharski[2], G. Gywiński[1,2], W. Knap[1,2], S. Rumyantsev[2]

[1]CENTERA, CEZAMAT, Warsaw University of Technology, 02-822 Warsaw, Poland

[2] Institute of High Pressure Physics PAS, 01-142 Warsaw, Poland

1. **Transmittance spectra of the plasmonic crystals at different temperatures and under UV illumination.**

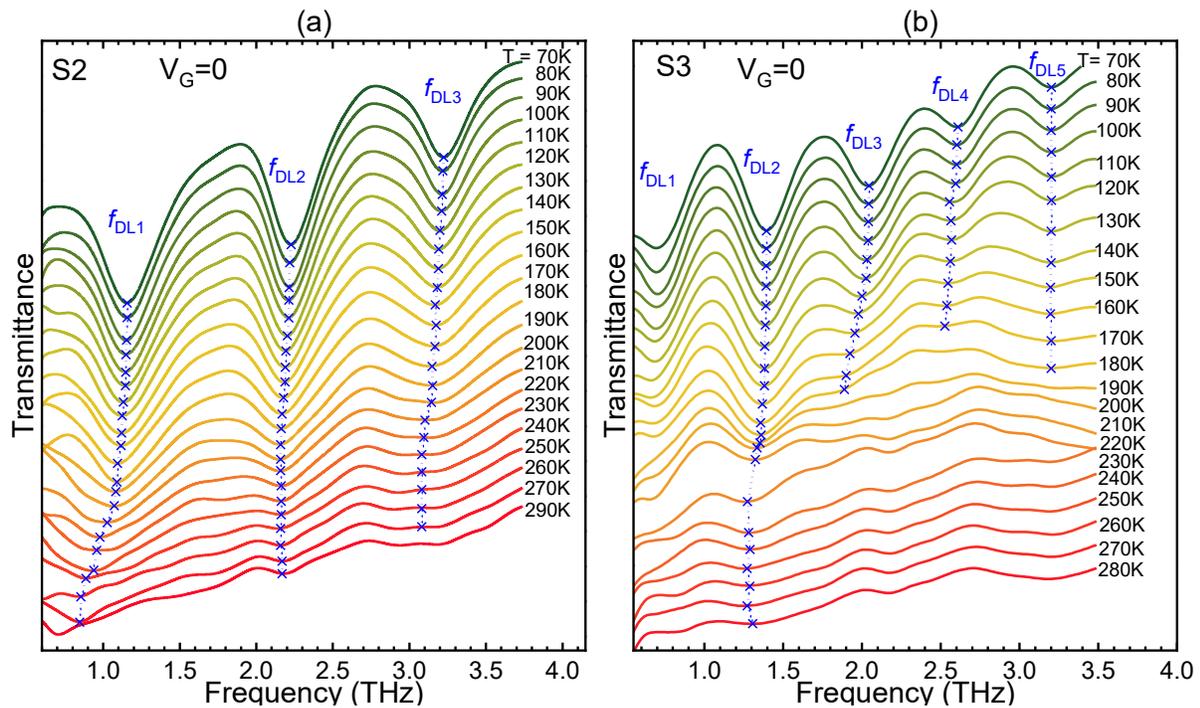

FIG. I. Transmittance spectra for S2 and S3 PC at different temperatures. Crosses show the position of the plasmon resonance frequencies. Dashed lines are guides for the eyes. The first harmonic of a delocalized plasmon in S3 PC is close to the low-frequency limit of the setup and cannot be used for the analysis.

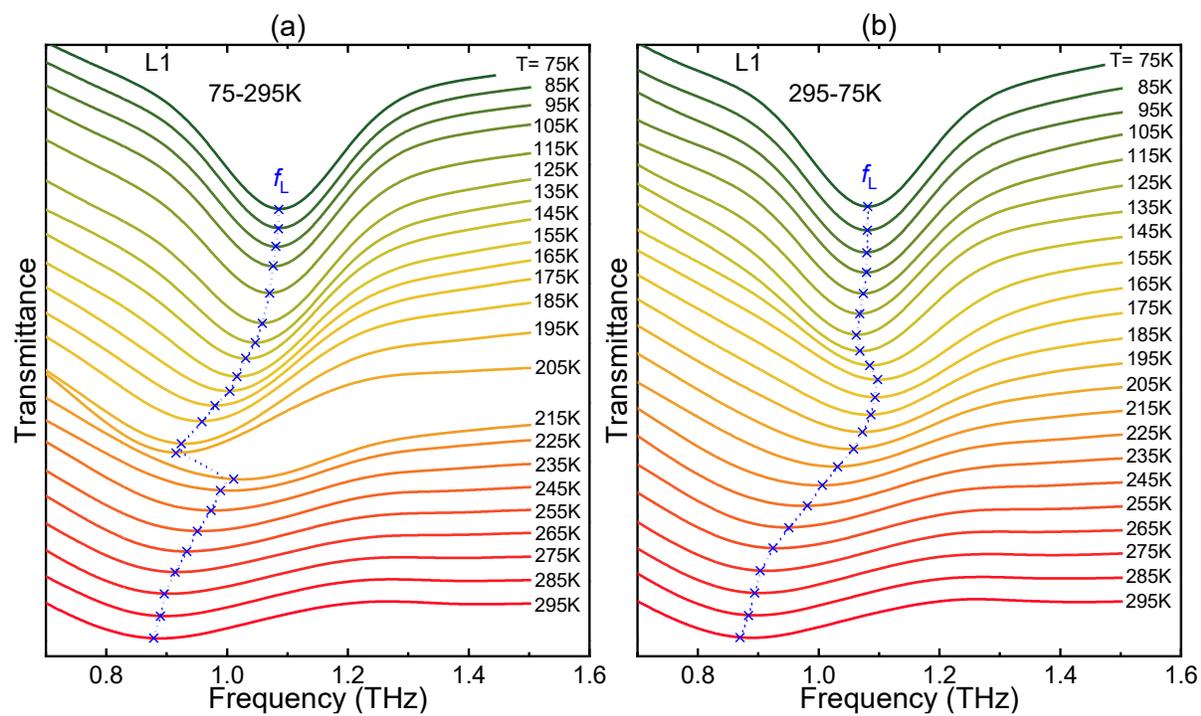

FIG. II. Transmittance spectra for the L1 PC at different temperatures during heating (a) and cooling (b) cycles. Crosses show the position of the plasmon resonance frequencies. Dashed lines are guides for eyes.

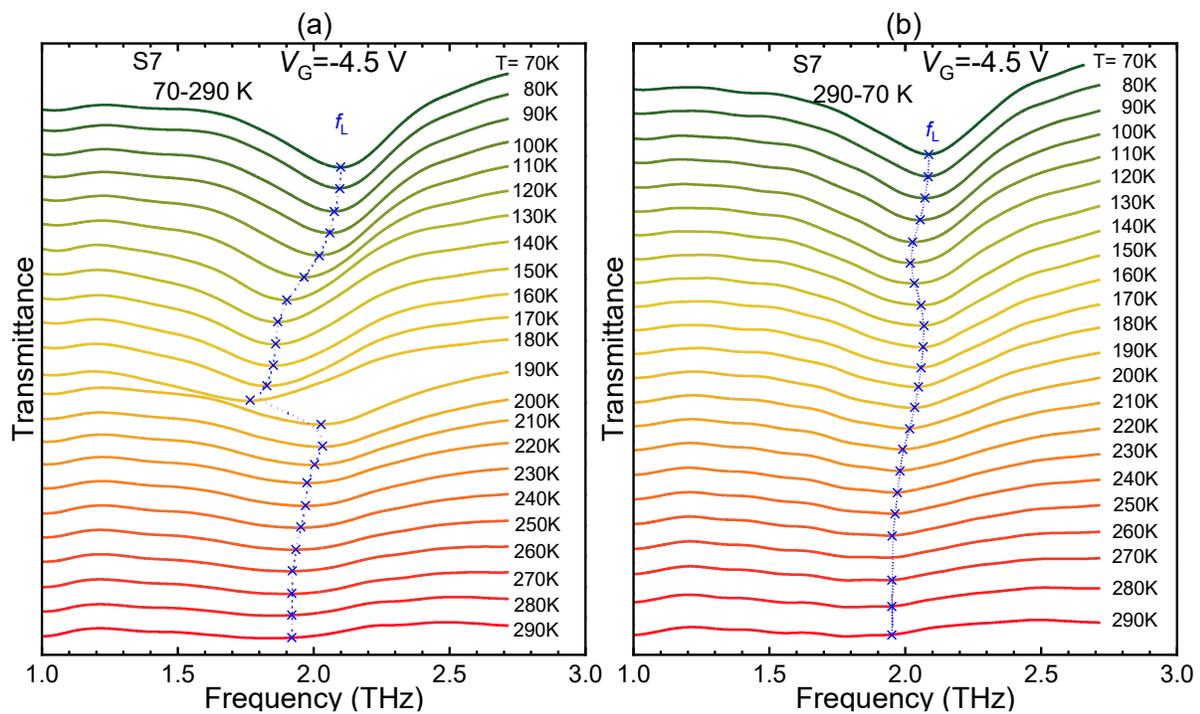

FIG. III. Transmittance spectra for the localized plasmons in S7 PC at different temperatures during heating (a) and cooling (b) cycles. Crosses show the position of the plasmon resonance frequencies. Dashed lines are guides for eyes.

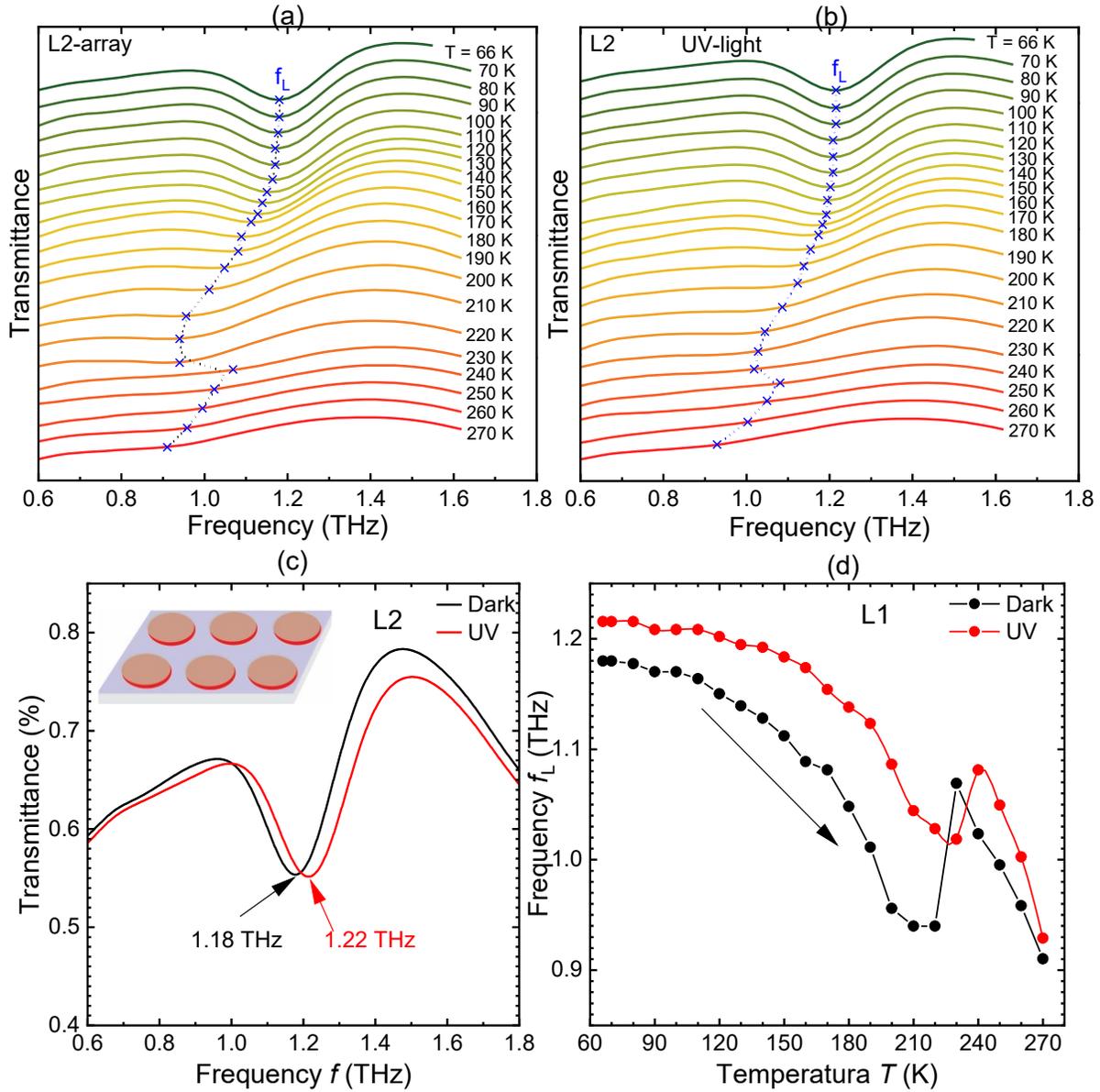

FIG. IV. Transmittance spectra for the L2 PC at different temperatures in dark (a) and under UV illumination (b). Crosses show the position of the plasmon resonance frequencies. Dashed lines are guides for eyes. Effect of the UV illumination on the plasmon spectrum transmittance spectrum for L2 PC. Temperature dependence of the resonant plasmon frequency in the dark and under UV illumination during the heating cycle (c) . Illumination wavelength and power are $\lambda = 270$ nm and $P \approx 50$ µW, respectively. (d)

## 2. Analysis of higher plasmon resonance harmonics

Several harmonics are usually observed on the transmittance spectra od PCs. Figure V shows the transmittance spectra for S1, S2, S3, S7 PCs at different gate voltages. The frequencies of higher harmonics also depend on the plasmon velocities in the gated and ungated parts of a PC. It is clearly seen that the frequency of every minima depends on the gate voltage, i.e. on the concentration in the gated parts of the PC. This is an additional confirmation that all minima in the transmittance spectra indeed belong to the first and higher harmonics of the plasmons.

Figure VI shows the temperature dependences of the same parameter $(f_{DL70}/f_{DL})^2$ as in Fig. 6 but for higher plasmon harmonics for several PCs (see Fig. I for the transmittance spectra of S2 and S3 PC at different temperatures). One can see that the dependences are weaker than those shown in Fig. 6. It was shown in ref. [4] that although the analytical model yields the correct frequency for the first harmonic of the plasmon resonance, it does not provide the correct plasmon frequency for higher harmonics and the higher is the harmonic frequency the higher is the disagreement.

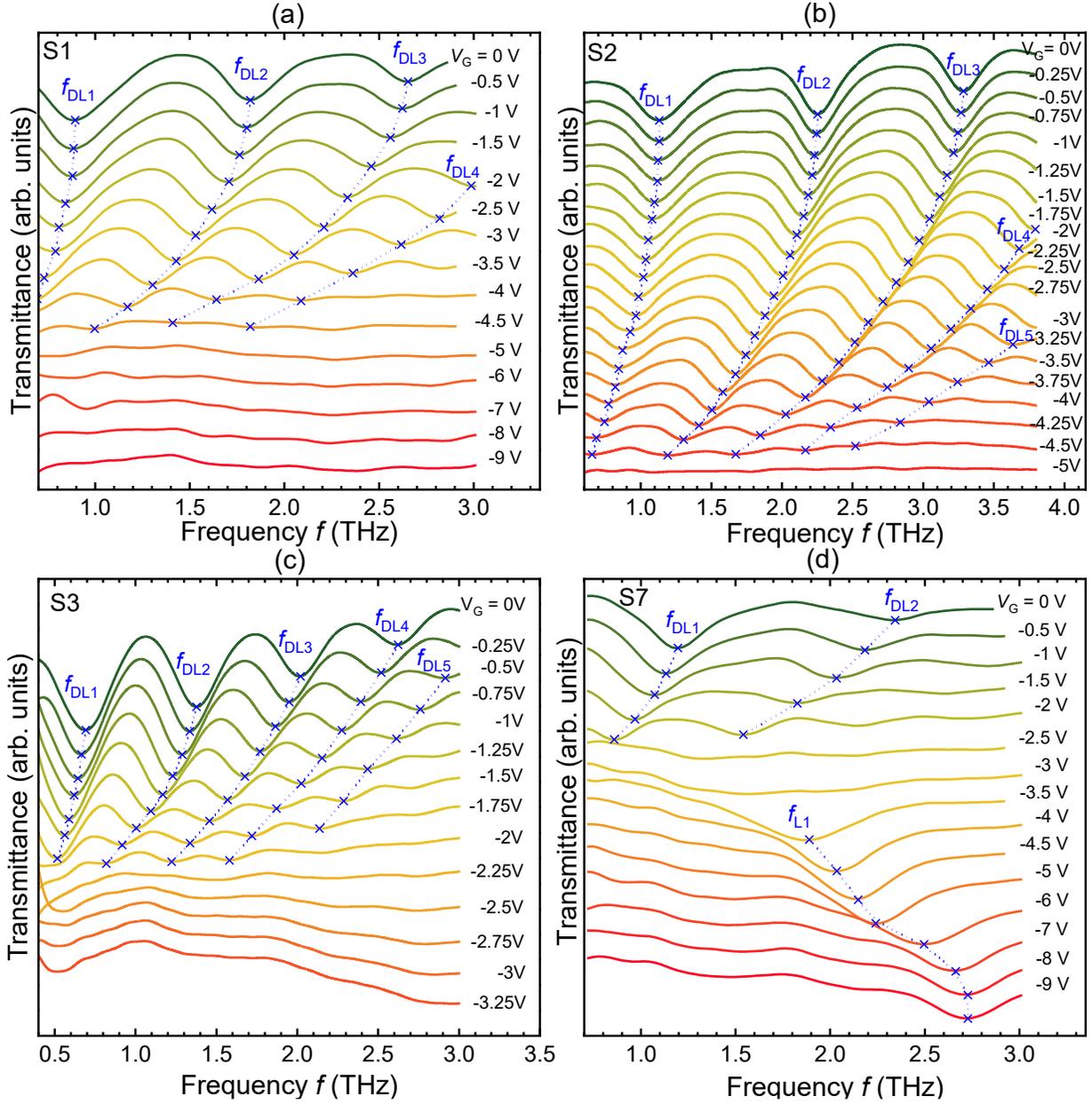

FIG. V. Transmittance spectra at different gate voltages, $V_G$, for S1, S2, S3, S7 PCs at temperature 70 K. All spectra are shown with vertical shifts for better visibility. Blue crosses show the position of the plasmon resonance frequencies. Dashed lines are guides for the eye.

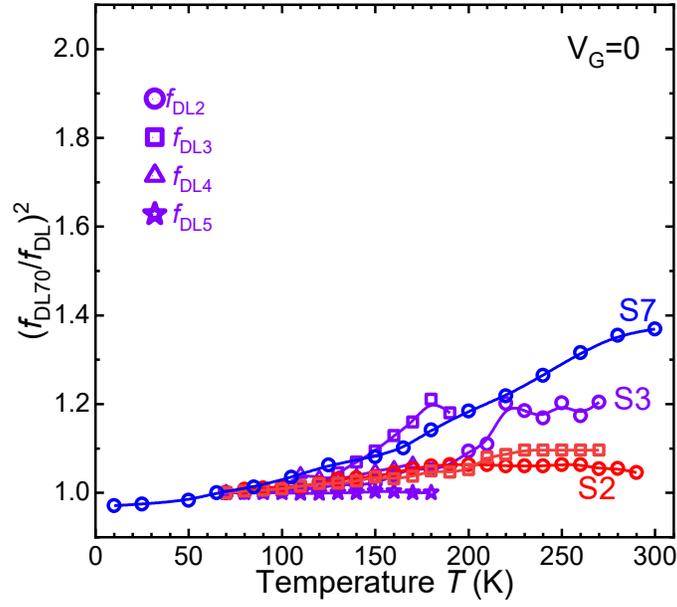

FIG. VI. Temperature dependence of the $(f_{DL70}/f_{DL})^2$ parameter for the higher harmonics of the delocalized mode of plasmon resonances for S2 (red), S3 (blue), and S7 ( black) PCs. Different symbols correspond to different harmonics as marked in the figure. Inset shows the transmittance spectra for the S1 PC. The gate voltage independent absorption line can be attributed to the $E_2$ optical phonon mode.